%% file: full_paper_main.tex
\documentclass{IET-Conf-Paper}

\usepackage[prependcaption,colorinlistoftodos]{todonotes}

\usepackage{xcolor}
\usepackage{amssymb,amsfonts}
\usepackage{graphicx}
\usepackage{textcomp}
\usepackage{booktabs}
\usepackage{multirow}
\usepackage{tabularx}
\usepackage{tablefootnote} 

\definecolor{PV}{rgb}{1, 0.75, 0}
\definecolor{wind}{rgb}{0.4, 0.8, 1}
\definecolor{STATCOM}{rgb}{0.6, 0.8, 0.3}

\header{22$^{nd}$ Wind \& Solar Integration Workshop | Copenhagen, Denmark | 26--28 September 2023}
\begin{document}

\title{EXPERIMENTAL VALIDATION OF A\\ DYNAMIC VIRTUAL POWER PLANT CONTROL\\ CONCEPT BASED ON A MULTI-CONVERTER\\ POWER HARDWARE-IN-THE-LOOP TEST BENCH}

\author{Moritz Andrejewski \ad{1}\corr, Verena Häberle \ad{2}, Nico Goldschmidt \ad{1},\\ Florian Dörfler \ad{2}, Horst Schulte \ad{1}}

\address{\add{1}{Control Engineering Group, University of Applied Sciences (HTW) Berlin, 
Faculty 1: School of Engineering -  Energy and Information, 12459 Berlin, Germany}
\add{2}{Automatic Control Laboratory, ETH Z\"urich, 8092 Z\"urich, Switzerland}
\email{moritz.andrejewski@HTW-Berlin.de}}


\begin{abstract}
Recently, the concept of dynamic virtual power plants (DVPP) has been proposed to collectively provide desired dynamic ancillary services such as fast frequency and voltage control by a heterogeneous ensemble of distributed energy resources (DER). This paper presents an experimental validation of a recent DVPP control design approach on a multi-converter power hardware-in-the-loop (PHIL) test bed system. More specifically, we consider a DVPP composed of a wind generation system, a photovoltaic (PV) system, and a STATCOM with small storage capacity to collectively provide grid-following fast frequency regulation in the presence of grid-frequency and load variations. The performance of the aggregated DVPP response is evaluated with respect to its ability to match a desired dynamic behavior while taking practical limitations of the individual DVPP units into account.
\end{abstract}
\maketitle

\section{Introduction}

In future power systems, non-synchronous, distributed energy resources (DER) will be increasingly demanded to provide dynamic ancillary services. This induces significant challenges to handle weather-volatile renewable energy sources and device limitations of individual DERs. Recently, the concept of dynamic virtual power plants (DVPP) has been proposed to tackle dynamic ancillary services provision by DERs \cite{marinescu2022dynamic,haeberle2021control,bjork2022dynamic,zhong2021coordinated}. DVPPs are ensembles of heterogeneous DERs (all with individual constraints) aggregated to collectively provide desired dynamic ancillary services such as fast frequency and voltage control. Namely, while none of the DERs can provide these services consistently across all power/energy levels or all-time scales, a sufficiently heterogeneous group of DERs can do so.

Among the existing (all-theoretical) control design methods for DVPPs available in the literature \cite{haeberle2021control,bjork2022dynamic,zhong2021coordinated}, a highly versatile and fundamental approach has been presented in \cite{haeberle2021control}. Here, various DER units in a DVPP, which are connected at one bus in the power system, are controlled so that their overall behavior corresponds to a desired dynamic I/O behavior specified as a desired transfer function. The approach is based on a divide-and-conquer strategy, which is composed of two main steps: First, the disaggregation of the desired dynamic behavior among the DVPP units using \textit{dynamic participation factors (DPF)} to obtain local desired behaviors while taking device-specific, possibly time-varying DER constraints (e.g., power/energy limitations, response times, etc.) into account. In the second step, a local feedback control is designed for each DVPP unit to achieve the desired local behavior.
\begin{figure}[b!]
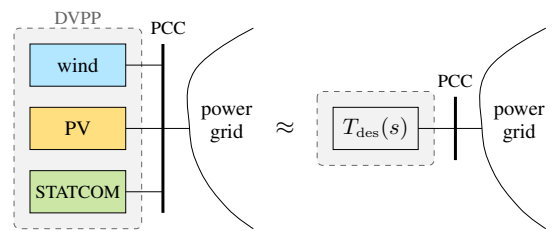

    \centering
    \vspace{-3mm}
    \include{Figures/DVPP_scheme}
    \vspace{-9mm}
    \caption{Sketch of a DVPP composed of a wind generation system, a PV generation system, and a STATCOM with small storage capacity to match a desired dynamic behavior $T_\mathrm{des}(s)$.}
    \label{fig:DVPP_scheme}
    \vspace{-6mm}
\end{figure}

This paper presents an experimental validation of the DVPP control concept in \cite{haeberle2021control} to provide grid-following fast frequency regulation. More specifically, we consider a DVPP composed of a wind generation system, a photovoltaic (PV) system, and a STATCOM with small storage (Fig. \ref{fig:DVPP_scheme}) to demonstrate and approve the effectiveness of the proposed DVPP control strategy in the presence of grid-frequency and load variations by means of a multi-converter power hardware-in-the-loop (PHIL) test bed system. In particular, in our experimental studies, we can verify the successful performance of the proposed DVPP concept, which turns out to be compliant with the simulation results in \cite{haeberle2021control}. In doing so, we particularly highlight the superiority of the employed DPFs over conventional static allocation schemes for frequency regulation, e.g., via droop gains. Namely, the use of DPFs allows taking (possibly time-varying) resource limitations in response time and capacity into account, while conventional droop schemes typically disregard and thus violate them, or alternatively require a vast over-dimensioning of the resource capacity to ensure a reliable operation.

The test bed system in our experimental validation setup contains three 11 kW back-to-back converter systems connected at a point of common coupling (PCC) and a 22 kW synchronous generator operated on a scaled AC grid with a load unit. The primary source technologies are emulated on an industrial programmable logic controller (PLC) from Bachmann, where the DVPP controller is implemented to generate the power setpoints for the converter systems as a function of the measured grid frequency deviation provided by a grid acquisition module. The synchronous generator provides a frequency-variable grid so that reactions of the DVPP to long-term changes in the grid frequency can be performed. Short-term grid frequency changes are generated via load increase or reduction via the load unit. For the validation of the DVPP, the converter systems are operated in the grid-following mode. During the experiments, the performance of the aggregated DVPP is evaluated with respect to its ability to provide a desired dynamic behavior for fast frequency regulation.

The remainder of this paper is structured as follows. In Section \ref{sec:dvpp_concept}, we recall the DVPP control concept in \cite{haeberle2021control}, tailored to the particular application that is experimentally validated in this work. We consider a simplified setup using the formalism of linearized systems, which makes it convenient to develop the control design. Section \ref{sec:testbed} introduces the PHIL experimental test bed system composed of multiple converter systems and the synchronous generator. In Section \ref{sec:case_studies}, we present our experimental validation results of the proposed DVPP control concept based on DPFs, where we particularly demonstrate its superiority over conventional static allocation strategies (e.g., via droop gains). Finally, Section \ref{sec:conclusion} summarizes the main results and discusses open questions.

\section{DVPP Control Concept}\label{sec:dvpp_concept}
In this section, we recall the DVPP control concept theoretically presented in \cite{haeberle2021control}, tailored to the particular application setup which is experimentally validated in this work.
\subsection{Control Setup}
\begin{figure}[t!]
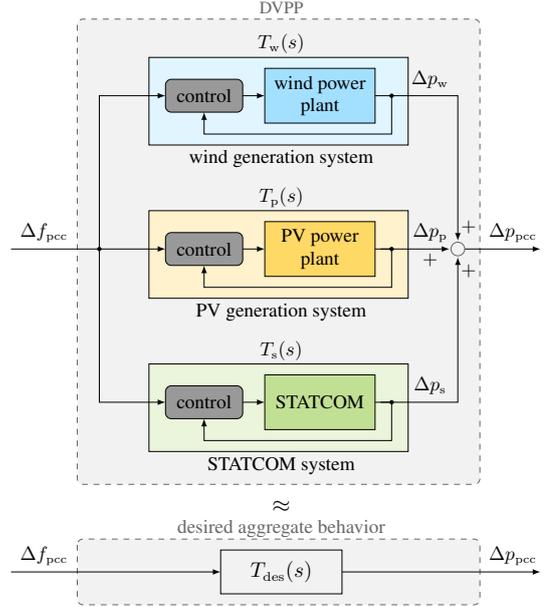

    \centering
    \vspace{-1mm}
    \include{Figures/DVPP_control_setup}
    \vspace{-9mm}
    \caption{Schematic of the DVPP control setup.}
    \vspace{-3mm}
    \label{fig:DVPP_control_setup}
\end{figure}

We consider a DVPP control setup for a group of heterogeneous DERs (Fig. \ref{fig:DVPP_control_setup} and Table \ref{tab:nomenclature}), including a wind generation system ($\mathrm{w}$), a PV generation system ($\mathrm{p}$), and a STATCOM system ($\mathrm{s}$). All DVPP units are connected to the power grid at the same point of common coupling (PCC) (Fig. \ref{fig:DVPP_scheme}), where they receive a common input signal in terms of the measured bus frequency deviation $\Delta f_\mathrm{pcc}$. The active power deviation outputs $\Delta p_i,\, i \in \{ \mathrm{w,p,s}\}$ of the wind, the PV, and the STATCOM (deviating from the respective power set point), sum up to the aggregate active power deviation output $\Delta p_\mathrm{pcc}$ at the PCC, i.e., 
\begin{align}\label{eq:aggregate_power}
    \Delta p_\mathrm{pcc} = \Delta p_\mathrm{w}+\Delta p_\mathrm{p}+\Delta p_\mathrm{s}.
\end{align}
When considering linearized systems, the local closed-loop dynamics of the DVPP units that map the measured frequency deviation $\Delta f_\mathrm{pcc}$ at the PCC to the local active power outputs $\Delta p_\mathrm{w}$, $\Delta p_\mathrm{p}$ and $\Delta p_\mathrm{s}$, respectively, can be described by the local closed-loop transfer functions $T_\mathrm{w}(s)$, $T_\mathrm{p}(s)$ and $T_\mathrm{s}(s)$, i.e., 
\begin{align}\label{eq:local_dynamics}
    \Delta p_i(s) &= T_i(s) \Delta f_\mathrm{pcc}(s),\quad \forall i \in \{\mathrm{w,p,s} \}.
\end{align}
Notice that the local closed-loop transfer functions $T_i(s)$ capture all dynamics underlying the decoupled ``active power loop'' of the grid-following DER power electronics architecture, which maps from frequency measurement to active power injection, i.e., the power converter dynamics, the filter dynamics, the grid-side converter control loops, the dc-side dynamics, and, most importantly, the dynamics of the primary source technology, i.e., the wind power plant, the PV power plant, and the STATCOM with the associated power plant controller (for detailed implementation aspects, see Section \ref{sec:testbed}). 

Considering the local closed-loop transfer functions $T_i(s)$ of the DERs, the aggregate DVPP behavior at the PCC is given as
\begin{align}\label{eq:aggregated_dynamics}
    \Delta p_\mathrm{pcc}(s) = \textstyle\sum_{i\in\{\mathrm{w,p,s}\}} T_i(s)\Delta f_\mathrm{pcc}(s).
\end{align}

To compensate for ancillary services conventionally provided by synchronous generators in transmission networks, a classical $\mathrm{f}$-$\mathrm{p}$ frequency control behavior is specified for the aggregate DVPP as a desired transfer function as
\begin{align}\label{eq:desired_specification}
    \Delta p_\mathrm{pcc}(s) = \underset{=\,T_\mathrm{des}(s)}{\underbrace{\frac{-D}{\tau s+1}}} \Delta f_\mathrm{pcc}(s),
\end{align}
where $D=6.5$ is the desired droop coefficient, and the denominator with $\tau=0.25$s is included to filter out high-frequency dynamics. Notice that we assume $T_\mathrm{des}(s)$ to be provided by the power system operator, who is encoding grid-code requirements in the form of the desired transfer function, e.g., in our case, a simple droop control with a first-order filter. Nevertheless, the DVPP control concept in \cite{haeberle2021control} is not limited to provide any kind of desired dynamic behavior, especially relevant for future grid-code specifications. 

Finally, by matching \eqref{eq:aggregated_dynamics} and \eqref{eq:desired_specification}, the DVPP control design problem is to find local controllers for the DVPP units $i\in\{w,p,s\}$, such that the following aggregation condition holds:
\begin{align}\label{eq:aggregation_condition}
    \textstyle\sum_{i\in\{\mathrm{w,p,s}\}} T_i(s) = T_\mathrm{des}(s),
\end{align}
as indicated in Fig. \ref{fig:DVPP_control_setup}. In doing so, it is important to ensure that practical limitations of the DERs, including time-scale constraints as well as (potentially time-varying) limits on power availability, are not exceeded during normal operation.

\subsection{Divide-and-Conquer Strategy}
To solve the previous DVPP control design problem in \eqref{eq:aggregation_condition}, the DVPP concept in \cite{haeberle2021control} is based on a divide-and-conquer strategy, composed of two steps:
\begin{enumerate}
    \item Disaggregate the desired behavior $T_\mathrm{des}(s)$ among the DVPP units using dynamic participation factors to obtain local desired behaviors.
    \item Design a local feedback control for each DVPP unit to optimally match the local desired behavior.
\end{enumerate}
\subsubsection{Disaggregation via DPFs}\label{subsubsec:Disaggregation_DPFs}
Given the aggregation condition in \eqref{eq:aggregation_condition}, we disaggregate the desired transfer function $T_\mathrm{des}(s)$ to the individual DVPP units $i$ as
\begin{align}\label{eq:disaggregation}
    T_\mathrm{des}(s) = \textstyle\sum_{i\in\{ \mathrm{w,p,s}\}} m_i(s)T_\mathrm{des}(s) =  \textstyle\sum_{i\in\{ \mathrm{w,p,s}\}} T_i(s),
\end{align}
where the transfer functions $m_i(s)$ are \textit{dynamic participation factors (DPF)}, required to satisfy the participation condition $\textstyle\sum_{i\in\{\mathrm{w,p,s} \}}m_i(s)=1$ with equality on the frequency range of interest.

\begin{figure}[t!]
    \centering
    \vspace{-2mm}
    \scalebox{0.5}{\includegraphics[]{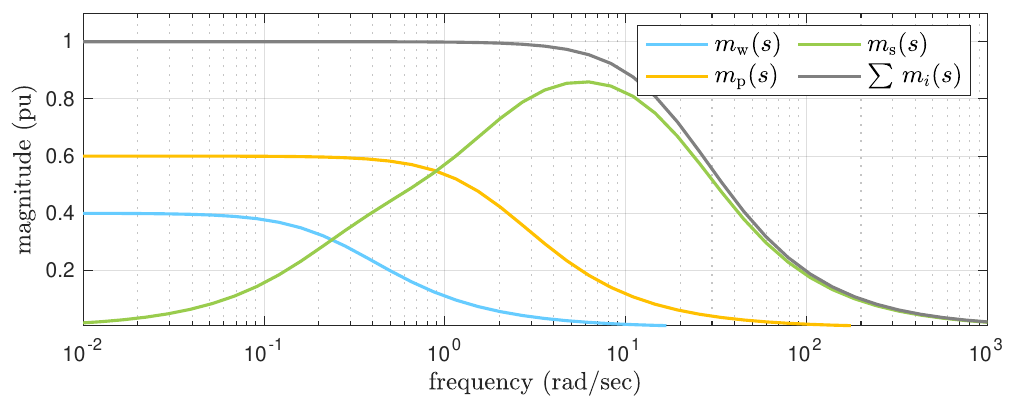}}
    \vspace{-4mm}
    \caption{Magnitude Bode plots of the selected DPFs for the wind, the PV and the STATCOM during nominal capacity conditions.}
    \vspace{-3mm}
    \label{fig:DPFs}
\end{figure}

The DPFs of the DVPP units are selected such that the previous participation condition is satisfied while simultaneously respecting the heterogeneous time scales of the local DER dynamics along with their steady-state power capacity limitations. In this regard, we select the DPFs for the wind and PV generation system according to a first-order low-pass filter participation behavior (Fig. \ref{fig:DPFs}) as
\begin{align}\label{eq:DPF_wind_pv}
        m_\mathrm{w}(s) = \frac{\mu_\mathrm{w}}{\tau_\mathrm{w}s+1},\quad\quad
        m_\mathrm{p}(s) = \frac{\mu_\mathrm{p}}{\tau_\mathrm{p}s+1},
\end{align}
where the time constants $\tau_\mathrm{w}=3.5$s and $\tau_\mathrm{p}=0.5$s for the roll-off frequency are selected according to the dominant dynamics of the wind and the PV power plant, respectively. Moreover, the low-pass filter dc gains $\mu_\mathrm{w}=0.4$ and $\mu_\mathrm{p}=0.6$ are chosen proportionately to the \textit{nominal} active power capacity limit of the wind and PV power plant while ensuring $\mu_\mathrm{w}+\mu_\mathrm{p}=1$. In order to \textit{approximately} satisfy the participation condition, the DPF of the STATCOM is specified to follow a band-pass filter behavior with high roll-off frequency $1/\tau_\mathrm{s}$ = $1/(0.05\text{s})$ (Fig. \ref{fig:DPFs}), intended to provide regulation on shorter time scales without steady-state contributions, i.e., 
\begin{align}\label{eq:DPF_statcom}
    m_\mathrm{s}(s) = \frac{1}{\tau_\mathrm{s}s+1}-m_\mathrm{w}(s)-m_\mathrm{p}(s).
\end{align}
Namely, by summing over the DPFs in \eqref{eq:DPF_wind_pv} and \eqref{eq:DPF_statcom}, we obtain $\textstyle\sum_{ i\in\{\mathrm{w,p,s}\} }m_i(s) = \tfrac{1}{\tau_\mathrm{s}s+1}\approx 1$, i.e., we tolerate a mismatch in the high-frequency range of the Bode plot of the overall DVPP response behavior.

As a special feature of the approach in \cite{haeberle2021control}, the low-pass filter dc gains $\mu_\mathrm{w}$ and $\mu_\mathrm{p}$ can be specified in such a way that they can be adapted online, in proportion to the possibly time-varying power capacity limits of the wind and PV power plant. If this is the case, we call $m_i(s)$ \textit{adaptive} DPFs, i.e., ADPFs, where $\mu_\mathrm{w}(t)$ and $\mu_\mathrm{p}(t)$ are now time-varying.

\subsubsection{Local Matching Control}
Finally, we need to find local feedback controllers for the DVPP units to ensure their local closed-loop transfer functions $T_i(s)$ match their local desired behavior, i.e., we impose the local matching condition
\begin{align}\label{eq:matching_cond}
    T_i(s) = m_i(s)T_\mathrm{des}(s),\quad \forall i \in \{ \mathrm{w,p,s}\}.
\end{align}
We resort to a proportional-integral (PI)-based matching control implementation, which we incorporate into the power plant control architecture of each DVPP unit (see Section \ref{sec:testbed}). Alternatively, more robust and optimal matching controllers can be obtained by using linear parameter-varying (LPV) $\mathcal{H}_\infty$ methods (see \cite{haeberle2021control}).

\renewcommand{\arraystretch}{0.9}
\begin{table}[t!]\footnotesize
    \centering
    \setlength{\tabcolsep}{1mm}
     \caption{List of notation for the DVPP control concept in \cite{haeberle2021control}.}
    \vspace{1mm}
    \begin{tabular}{c||c}
     \toprule
         Description &Symbol  \\ \hline
         Wind generation system index & $\mathrm{w}$ \\
         PV generation system index & $\mathrm{p}$ \\
         STATCOM system index & $\mathrm{s}$\\
         DVPP unit index $i\in\{ \mathrm{w,p,s}\}$ & $i$\\ \hline
         Measured bus frequency deviation at the PCC & $\Delta f_\mathrm{pcc}$\\
         Local active power deviation output of unit $i$ & $\Delta p_i$\\
         Local closed-loop transfer function of unit $i$ & $T_i(s)$\\
         Desired DVPP transfer function for $\mathrm{f}$-$\mathrm{p}$ control & $T_\mathrm{des}(s)$\\
         Desired droop coefficient & $D$\\
         Time constant for desired droop control & $\tau$ \\\hline
         DPF of unit $i$ & $m_i(s)$ \\
         (Possibly time-varying) dc gain of unit $i$ & $\mu_i$\\
         Time constant for the roll-off frequency of unit $i$ & $\tau_i$ \\
    \bottomrule
    \end{tabular}
\label{tab:nomenclature}
\end{table}
\renewcommand{\arraystretch}{1} \normalsize

\section{Multi-Converter PHIL Test Bed System}\label{sec:testbed}
\subsection{Simulation Environment}
\begin{figure}[t!]
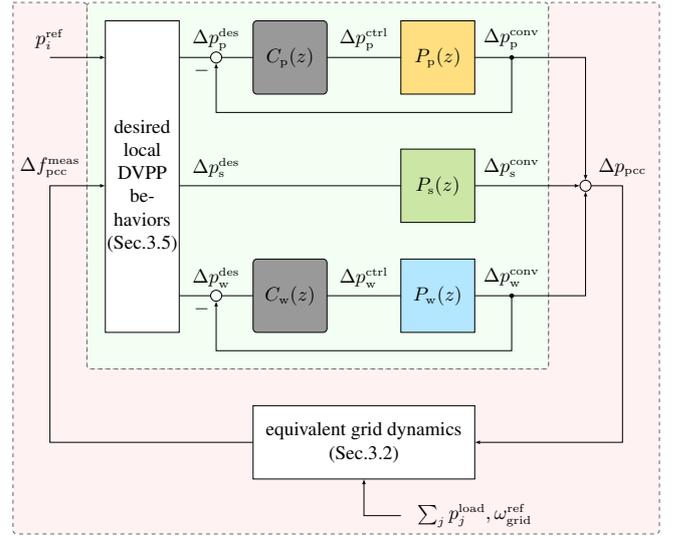

    \centering
    \include{Figures/DVPP_TF_scheme}
    \vspace{-10mm}
    \caption{Sketch of the DVPP structure in simulation.}
    \vspace{-2.5mm}
    \label{fig:DVPP_TF_scheme}
\end{figure}

In preparation for the experimental validation, the entire system and its behavior are designed in a discrete simulation environment with a constant task tick rate resp. sample time $T = t_{k+1} - t_{k}$ via Matlab 2019b with Simulink, and iteratively optimized to the behavior of the test bed system. 
All signals in the discrete simulation are sampled signals with the label $[\cdot](kT)$ where $t_k = kT$, which will be written simply as $[\cdot](k)$ in the following. The designed structure in Fig.~\ref{fig:DVPP_TF_scheme} illustrates the simplified closed-loop system, separated into two parts: The green-colored area is the part uploaded to a Bachmann MC220 programmable logic controller (PLC) via Autocode, while the red colored area includes the equivalent grid behavior to represent the test bed system of the experimental study.
\relax
\subsection{Equivalent Grid Equation}\label{subsec:equi_grid_eq}
The swing equation \eqref{eq:equi_grid_Lap} below represents the equivalent grid in discrete time. This results from the forward Euler method 
\begin{align}
\dot\omega^\mathrm{meas}_\mathrm{pcc}(k) 
\approx  \frac{\omega^\mathrm{meas}_\mathrm{pcc}(k+1) - \omega^\mathrm{meas}_\mathrm{pcc}(k)}{T} = 
   \frac{\omega^\mathrm{meas}_\mathrm{pcc}(k) (z-1)}{T}
\end{align}
with the shift operator $z$ where $y(k+1) = y(k) z$. The resulting change of the angular frequency $\omega^\mathrm{meas}_\mathrm{pcc}(k)$ as a function of the grid differential power $\Delta p_\mathrm{grid}(k)$ shows the swing behavior of the grid considering the reference angular frequency $\omega^\mathrm{ref}_\mathrm{grid}$. A desired grid configuration is possible with the inertia time constant $H$ and the damping ratio $d$, which, in this study, represents the behavior of the test bed system. The dynamics of $\omega^\mathrm{meas}_\mathrm{pcc}(k)$ in \eqref{eq:equi_grid_Lap} are influenced by two parameters: The input variable $\Delta p_\mathrm{grid}(k)$ and the constant parameter $\omega^\mathrm{ref}_\mathrm{grid}$ representing a disturbance in the transfer function:
\vspace{-0.25cm}
\begin{align}\label{eq:equi_grid_Lap}
\begin{split}
    \omega^\mathrm{meas}_\mathrm{pcc}(k) & = \frac{\frac{1}{2\,H}}{\left(\frac{z - 1}{T}\right) + \frac{d}{2\,H}}\underbrace{\left[\Delta p_\mathrm{grid}(k) 
    + d\,\omega^\mathrm{ref}_\mathrm{grid}\right]}_{\Delta p_{d}(k)}.
    \end{split}
\end{align}
By substituting in $\Delta p_{d}(k)$, the time-discrete transfer function of the grid model follows as
\vspace{-0.5cm}
\begin{align}
    \begin{array}{l}
        \begin{array}{rcl}
    P_\mathrm{grid}(z)    & = & \dfrac{\omega^\mathrm{meas}_\mathrm{pcc}(k)}{\Delta p_\mathrm{d}(k)}.
        \end{array} 
    \end{array}
\end{align}
Based on the latter, the resulting frequency difference $ \Delta f^\mathrm{meas}_\mathrm{pcc}(k)$ of the grid relative to the constant reference frequency $f^\mathrm{ref}_\mathrm{grid} = \frac{\omega^\mathrm{ref}_\mathrm{grid}}{2\pi}$, which is usually measured at the PCC, can be obtained as
\vspace{-0.35cm}
\begin{align}\label{eq:delta_f_pcc}
    \Delta f^\mathrm{meas}_\mathrm{pcc}(k) = \dfrac{1}{2\pi}\left(\omega^\mathrm{meas}_\mathrm{pcc}(k) - \omega^\mathrm{ref}_\mathrm{grid}\right). 
\end{align}
The change in frequency is primarily dependent on the balance of the generated power $p^\mathrm{conv}_{i}(k)$ to the consumed power $p^\mathrm{load}_{j}(k)$ of all grid participants, where
\vspace{-0.25cm}
\begin{align}\label{eq:delta_P_g}
    \Delta p_\mathrm{grid}(k) = \textstyle\sum_{i\in\{ \mathrm{w,p,s}\}} p^\mathrm{conv}_{i}(k) - \textstyle\sum_{j} p^\mathrm{load}_{j}(k). 
\end{align}

\renewcommand{\arraystretch}{0.9}
\begin{table}[t!]\footnotesize
    \centering
    \setlength{\tabcolsep}{1mm}
     \caption{List extension for the discrete simulation environment.}
    \vspace{1mm}
    \begin{tabular}{c||c}
     \toprule
        Description                                                                         & Symbol  \\        \hline
        Normally measured value in test bed system                                          & $\mathrm{meas}$\\
        Set power value for units in test bed system                                       & $p^\mathrm{conv}_{i}(k)$\\
        Total active power capacities of the unit $i$                                       & $p^\mathrm{ref}_{i}(k)$\\
        Load power value of load units in test bed system                                                       & $p^\mathrm{load}_{j}(k)$\\
        Grid value in test bed system                                                       & $\mathrm{grid}$\\
        Control value of matching controller                                                 & $\mathrm{ctrl}$\\\hline
        Load unit index $j$                                                                 & $j$\\\hline
        Local active power deviation output of unit $i$                                     & $\Delta p_i(k)$\\
        Aggregate active power deviation output at the PCC                                  & $\Delta p_\mathrm{pcc}(k)$\\\hline
        Inertia time constant of the equivalent grid                                        & $H$\\
        Damping ratio of angular frequency change of equivalent grid                        & $d$\\             \hline
        time-discrete transfer function of plant $i$                                        & $P_i(z)$\\
        time-discrete transfer function of matching controller $i$                          & $C_i(z)$\\
    \bottomrule
    \end{tabular}
        \vspace{-3mm}
\label{tab:nomenclature_discrete}
\end{table}
\renewcommand{\arraystretch}{1} \normalsize

\subsection{Primary Source Characteristic}\label{subsec:prim_source_char}
The test bed system consists of multiple converters as generating units, where a transfer function representation of the underlying primary source, i.e., of the wind power plant, the PV system, and the STATCOM, is additionally included. 
The transfer functions for PV and wind are designed as nonlinear closed-loop systems as in \cite{AlbernazLacerda2022}, and can be linearized with a defined operating case to achieve simplified behaviors.

The internally controlled PV power plant is used to track the maximum power point (MPP) or demand power point (DPP) by applying a perturb and observe (P\&O) method \cite{AlbernazLacerda2022}.  The feedback information of the linearized closed-loop PV system is the electrical power of the PV converter $p^\mathrm{conv}_\mathrm{p}(k)$, and the manipulated variable $p^\mathrm{ctrl}_\mathrm{p}(k)$ corresponds to the reference power of the closed-loop PV system. The associated linearized time-discrete transfer function is given as
\vspace{-0.25cm}
\begin{align}
    \begin{array}{l}
        \begin{array}{rcl}
    P_\mathrm{p}(z)    & = & \dfrac{p^\mathrm{conv}_\mathrm{p}(k)}{p^\mathrm{ctrl}_\mathrm{p}(k)}
        \end{array}    \\[1.1em]
        \tiny
        \begin{array}{rcl}
                & = & \dfrac{a_\mathrm{p,4} z^{4} + a_\mathrm{p,3} z^{3} + a_\mathrm{p,2} z^{2} + a_\mathrm{p,1} z + a_\mathrm{p,0}}{b_\mathrm{p,5} z^{5} + b_\mathrm{p,4} z^{4} + b_\mathrm{p,3} z^{3} + b_\mathrm{p,2} z^{2} + b_\mathrm{p,1} z + b_\mathrm{p,0}},    
        \end{array}
    \end{array}
\end{align}
with the operating-point dependent coefficients    
\vspace{-0.25cm}
\begin{align}
    \scriptsize
    \begin{array}{rcl}
        a_\mathrm{p,4} & = & 0.4028    \\[0.2em]
        a_\mathrm{p,3} & = & -1.0303   \\[0.2em]
        a_\mathrm{p,2} & = & 1.0041    \\[0.2em]
        a_\mathrm{p,1} & = & -0.3767   \\[0.2em]
        a_\mathrm{p,0} & = & 8.4638 \cdot 10^{-5} 
    \end{array}
    \begin{array}{rcl}
        b_\mathrm{p,5} & = & 1         \\[0.2em]
        b_\mathrm{p,4} & = & -2.3955   \\[0.2em]
        b_\mathrm{p,3} & = & 2.0413    \\[0.2em]
        b_\mathrm{p,2} & = & -0.7444   \\[0.2em]
        b_\mathrm{p,1} & = & 0.0985    \\[0.2em]
        b_\mathrm{p,0} & = & -3.575 \cdot 10^{-9} \textnormal{.}
    \end{array}
\end{align}

Also, the wind power plant is modeled as a closed-loop system, linearized for the operation point of wind turbine type 4, the full converter variant in full-load operation. Load reduction is achieved via pitch adjustment. The system is internally controlled via a Takaki-Sugeno modeling framework that uses a convex description of linear models to describe nonlinear dynamics, with a focus on mechanical load reduction and perfect control of the turbine states
\cite{Poeschke2020}, \cite{Poeschke2021}. The feedback information of linearized closed-loop wind power system is the electrical output power of the generator $p^\mathrm{conv}_\mathrm{w}(k)$, and the manipulated variable $p^\mathrm{ctrl}_\mathrm{w}(k)$ corresponds to the reference power of the wind power plant. The associated linearized transfer function is given as
\vspace{-0.25cm}
\begin{align}\label{eq:TF_wind}
    \begin{array}{l}
        \begin{array}{rcl}
            P_\mathrm{w}(z)    & = & \dfrac{p^\mathrm{conv}_\mathrm{w}(k)}{p^\mathrm{ctrl}_\mathrm{w}(k)}
        \end{array}     \\[1.2em]
        \tiny\hspace{-0.6cm}
        \begin{array}{rcl}
                    & = & \dfrac{a_\mathrm{w,6} z^{6} + a_\mathrm{w,5} z^{5} + a_\mathrm{w,4} z^{4} + a_\mathrm{w,3} z^{3} + a_\mathrm{w,2} z^{2} + a_\mathrm{w,1} z + a_\mathrm{w,0}}{b_\mathrm{w,7} z^{7} + b_\mathrm{w,6} z^{6} + b_\mathrm{w,5} z^{5} + b_\mathrm{w,4} z^{4} + b_\mathrm{w,3} z^{3} + b_\mathrm{w,2} z^{2} + b_\mathrm{w,1} z + b_\mathrm{w,0}},    
        \end{array}
    \end{array}
\end{align}
with the specific coefficients, depending on the selected operating point of the produced power generated by the current wind speed, which is shown as follows
\vspace{-0.25cm}
\begin{align}\label{eq:TF_wind_coeff}
    \scriptsize
    \begin{array}{rcl}
        a_\mathrm{w,6} & = & 0.1943    \\[0.2em]
        a_\mathrm{w,5} & = & -1.1346   \\[0.2em]
        a_\mathrm{w,4} & = & 2.7637    \\[0.2em]
        a_\mathrm{w,3} & = & -3.5947   \\[0.2em]
        a_\mathrm{w,2} & = & 2.6328    \\[0.2em]
        a_\mathrm{w,1} & = & -1.0295   \\[0.2em]
        a_\mathrm{w,0} & = & 0.01679 
    \end{array} 
    \begin{array}{rcl}
        b_\mathrm{w,7} & = & 1         \\[0.2em]
        b_\mathrm{w,6} & = & -6.4395   \\[0.2em]
        b_\mathrm{w,5} & = & 17.7314   \\[0.2em]
        b_\mathrm{w,4} & = & -27.0496  \\[0.2em]
        b_\mathrm{w,3} & = & 24.6770   \\[0.2em]
        b_\mathrm{w,2} & = & -13.4537  \\[0.2em]
        b_\mathrm{w,1} & = & 4.0553    \\[0.2em]
        b_\mathrm{w,0} & = & -0.5208 \textnormal{.}
    \end{array}
\end{align}
No additional dynamics are imprinted on the STATCOM storage. Rather, it is assumed that the time delay of the power transmission of the STATCOM storage in relation to the other components is much smaller and therefore results in an ideal transmission. The goal is to reach a contrast to the other DVPP units with the very fast response of the STATCOM to achieve a more heterogeneous ensemble which illustratively highlights the strengths of a DVPP. The feedback information is the sampled electrical output power $p^\mathrm{conv}_\mathrm{s}(k)$ which is equal to the desired variable of DVPP $p^\mathrm{des}_\mathrm{s}(k)$. The transfer function is given as 
\vspace{-0.15cm}
\begin{align}\label{eq:TF_STATCOM}
    P_\mathrm{s}(z)    & = \dfrac{p^\mathrm{conv}_\mathrm{s}(k)}{p^\mathrm{des}_\mathrm{s}(k)} = 1.
\end{align}

Finally, the reference value for the aggregates sent to the converters is the total value $p^\mathrm{conv}_\mathrm{i}(k)$, related to the reference power of the aggregate $p^\mathrm{ref}_\mathrm{i}(k)$ with
\vspace{-0.15cm}
\begin{align}
    p^\mathrm{conv}_{i}(k) = \Delta p^\mathrm{conv}_{i}(k) + p^\mathrm{ref}_{i}(k), \quad \forall i\in\{\mathrm{w,p,s}\}.
\end{align}
\subsection{Matching Control Structure}\label{subsec:match_ctrl_struct}
 For optimal achievement of the desired power matching by the DVPP, the wind and PV generation systems are equipped with higher-level matching controllers. Given the ideal tracking characteristic of the considered STATCOM storage in \eqref{eq:TF_STATCOM}, we do not need an additional matching controller in the case of the latter. For the matching controllers of the wind and PV systems, we resort to discrete-time PID controllers denoted as $C_\mathrm{p}(z)$ and $C_\mathrm{w}(z)$ in Fig.~\ref{fig:DVPP_TF_scheme}. The associated control gains of each controller $i\in\{\mathrm{w,p}\}$ are designed separately and optimized for the behavior in the test system. 
The control error $e_{i}(k)$ results from the desired variable $p^\mathrm{des}_{i}(k)$ and the feedback variable $p^\mathrm{conv}_{i}(k)$ of each DVPP unit $i$, and follows accordingly as
\vspace{-0.25cm}
\begin{align}\label{eq:PID_control_Diff}
    e_{i}(k) = p^\mathrm{des}_{i}(k) - p^\mathrm{conv}_{i}(k),\quad i\in\{\mathrm{w,p}\}\textnormal{.}
\end{align}

A special extension is the saturation limit of the controller output, which is dynamically adjusted in relation to the possibly time-varying power limits of the wind and PV plants.

\subsection{Desired Power of the local DVPP Reference Models}\label{subsec:DPF_struct}
As described in Section \ref{subsubsec:Disaggregation_DPFs}, the difference of desired active power injections for each aggregate $\Delta p^\mathrm{des}_i(k)$ are obtained from the desired time-discrete transfer function $T_\mathrm{des}(z)$ and the selected DPF $m_{i}(z),\,\forall i\in\{\mathrm{w,p,s}\}$ as in \eqref{eq:matching_cond}. In this regard, for the wind and PV system $i\in\{\mathrm{w,p}\}$, the time-discrete form of the desired local behavior is accordingly given as 
\vspace{-0.15cm}
\begin{align}
 m_{i}(z) T_\mathrm{des}(z) = \left(\frac{\mu_{i}}{\tau_{i}\left(\frac{z - 1}{T }\right) + 1}\right) \left(\frac{-D}{\tau \left(\frac{z - 1}{T}\right) + 1}\right),
\end{align}
with the proportionality dc gain specified as  
\vspace{-0.15cm}
\begin{align}
    \mu_{i} = \dfrac{p^\mathrm{ref}_{i}(k)}{\textstyle\sum_{i\in\{ \mathrm{w,p}\}} p^\mathrm{ref}_{i}(k)}    \textnormal{.}
\end{align}
This results in the desired local active power injections of the wind and PV system as 
\begin{align}
    \Delta p^\mathrm{des}_i(k) & = m_i(z) T_\mathrm{des}(z)  \, \Delta f^\mathrm{meas}_\mathrm{pcc}(k),\quad i\in\{\mathrm{w,p}\}.
\end{align}
Given the latter, the desired local active power injection of the STATCOM is finally obtained in accordance with  \eqref{eq:DPF_statcom} as 
\begin{align}
\begin{split}
    \Delta p^\mathrm{des}_{s}(k) =& \dfrac{1}{\tau_{s} \left(\frac{z - 1}{T}\right) + 1} T_\mathrm{des}(z) \Delta f^\mathrm{meas}_\mathrm{pcc}(k) \\
    &- \textstyle\sum_{i\in\{\mathrm{w,p}\}}\Delta p^\mathrm{des}_i(k) \textnormal{.}
    \end{split}
\end{align}


\subsection{Experimental Test Bed System}\label{sec:exp_testbed_system}
\begin{figure}[t!]
    \centering
    \scalebox{0.94}{\includegraphics[]{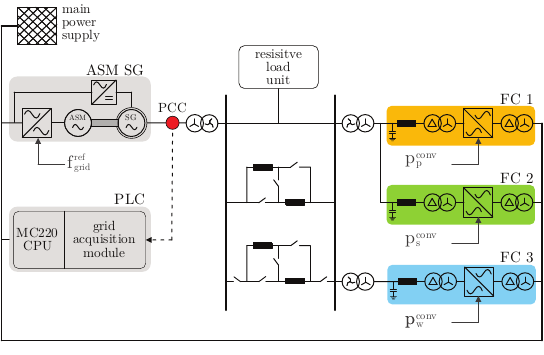}}
    \vspace{-4mm}
    \caption{Reduced single-pole schematic of the multi-converter PHIL test bed with PLC and signal lines for DVPP evaluation.}
    \label{fig:DVPP_Testbed_Schematic}
        \vspace{-3mm}
\end{figure}

\begin{figure}[t!]
    \centering
    \scalebox{0.88}{\includegraphics[]{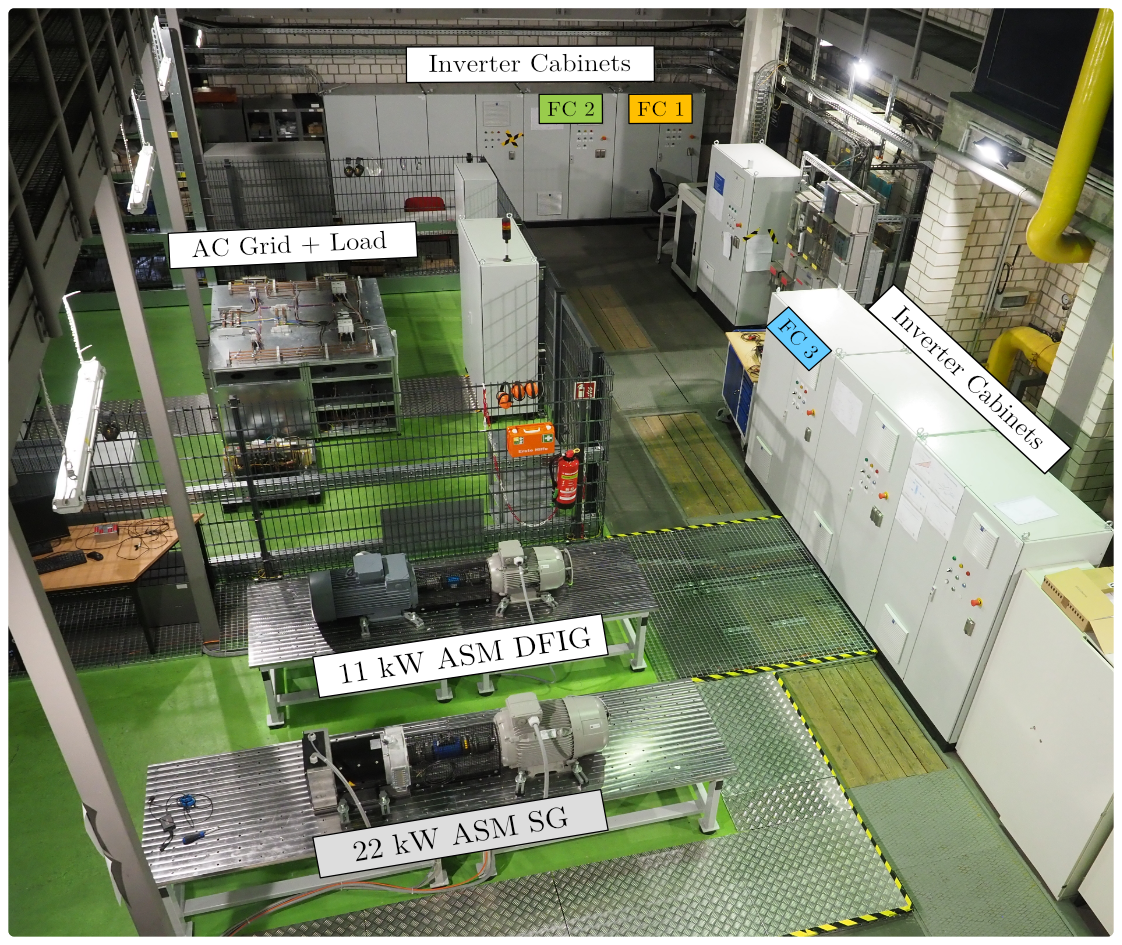}}
    \vspace{-3mm}
    \caption{Picture of the real test bed.}
    \label{fig:DVPP_Testbed_Picture}
      \vspace{-3mm}
\end{figure}

The experimental validation of the DVPP is performed at the multi-converter PHIL test bed of HTW-Berlin. 
Fig. \ref{fig:DVPP_Testbed_Schematic} shows a reduced single-pole schematic of the test bed, where only the hardware used for the DVPP tests is shown. Fig. \ref{fig:DVPP_Testbed_Picture} gives a real-world overview of the test bed.

The test bed offers the possibility to operate the DVPP concepts in a realistic environment. For this purpose, three 11 kW back-to-back frequency converter (FC) systems (FC 1, FC 2, FC 3) and a 22 kW synchronous generator (ASM SG) are operated on a scaled AC grid with a resistive load unit. The DVPP is executed on a Bachmann MC220 PLC and generates power demands for the three converter systems from the measured grid frequency at a PCC which is provided by a grid acquisition module. The converter systems are operated in the current controlled grid following mode. The synchronous generator provides a frequency-variable grid so that reactions of the DVPP to long-term changes in the grid frequency can be performed by adjustments to the reference frequency $f_\mathrm{grid}^\mathrm{ref}$. Short-term changes of the grid frequency are generated via the load unit by means of stepwise load increase or load reduction. The SG is driven by a PI-speed-controlled asynchronous machine (ASM), to set and match the required grid frequency $f_\mathrm{grid}^\mathrm{ref}$. The speed control can be switched between a "standard" and a "slow" mode which provides different control responses to load changes. Due to the PI-controlled speed regulation, the dynamics of the grid frequency is dominated by the ASM. This limits the influence of the active power feed-in generated by the DVPP via the converters on the behavior of the grid frequency.

\begin{figure}[t!]
    \centering
    \vspace{-3mm}
    \scalebox{0.88}{\includegraphics[]{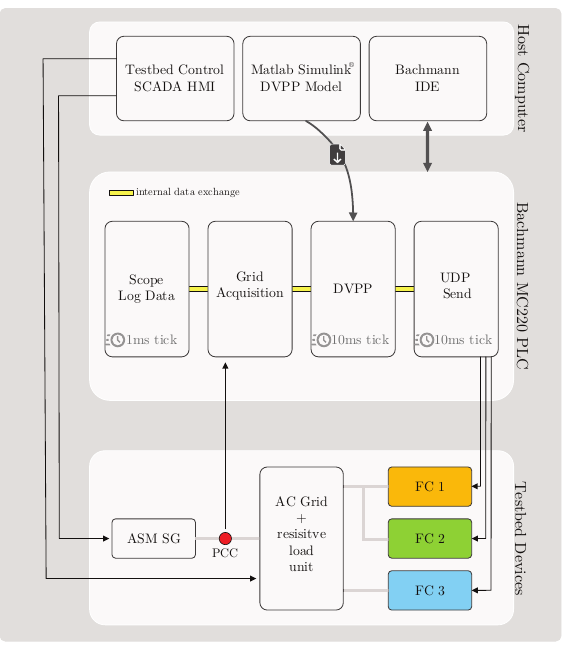}}
    \vspace{-3mm}
    \caption{Schematic of the DVPP testing process.}
    \label{fig:DVPP_Testbed_Workflow}
        \vspace{-3mm}
\end{figure}

The workflow for validating the DVPP at the test bed is shown in Fig. \ref{fig:DVPP_Testbed_Workflow}. A human-machine interface (HMI) at the host computer is used to parameterize the test bed. Therefore, the desired grid frequency and the speed control mode are demanded by the ASM and the desired load value is set to the resistive load unit. Matlab Simulink is used to compile and download the DVPP model with its structure shown in Fig. \ref{fig:DVPP_Testbed_Workflow} to the MC220 PLC target. The Bachmann IDE is used to control the processes and the datalogging on the PLC. Once the system setup is done, the operation of the DVPP running on the realtime target is enabled. To finally realize the DVPP testing there are three auxiliary task running on the PLC.
The actual grid frequency at PCC is provided to the DVPP by the Grid Acquisition Task. The active power values calculated by the DVPP are sent as power setpoints ($p_\mathrm{w}^\mathrm{conv}$, $p_\mathrm{p}^\mathrm{conv}$, $p_\mathrm{s}^\mathrm{conv}$) to the converter systems via the UDP Send task, where UDP means that the user datagram protocol is used for data transmission. The Scope Log Data Task is used to log all relevant data with a resolution of 1 kHz.

The initial conditions and operating parameters of the test bed system applied for the experimental case studies are provided in Section \ref{sec:testbed_para} and Table \ref{tab:testbed_para}.

\section{Simulational and Experimental Case Studies}\label{sec:case_studies}
To experimentally validate the proposed DVPP control concept, we perform two case studies on the multi-converter PHIL test bed system. In the \textit{first} case study, we investigate the long-term DVPP response behavior during a frequency jump of the grid frequency, where we particularly demonstrate the benefit of the DPFs over conventional static allocation schemes similar to droop control. In the \textit{second} case study, the DVPP is exposed to a short-term change of the grid frequency caused by a load jump of the load unit, where we study the effect of the DVPP on the resulting frequency response dynamics. 

\subsection{Test bed Parameters and Initial Conditions}\label{sec:testbed_para}
For the performance and reproducibility of the DVPP experiments, the default settings for the test bed and the equipment used were set as shown in Table \ref{tab:testbed_para}.
This initial operating state of the system serves as the basis for all test scenarios. Note that the frequency converters are operated in the current controlled grid following mode and the ASM SG is PI-speed-controlled as mentioned in Section \ref{sec:exp_testbed_system}. The grid voltage at PCC is provided by a voltage-controlled SG.
\renewcommand{\arraystretch}{0.9}
\begin{table}[ht!]\footnotesize
    \vspace*{-0.4cm}
    \centering
    \setlength{\tabcolsep}{1mm}
     \caption{List of test bed parameters and initial conditions.}
    \vspace{1mm}
    \begin{minipage}{0.5\textwidth}
    \begin{tabular}{c|c|c|c}
     \toprule
         Description &Component &Value & Value pu  \\ \hline
         Initial base load & Load Unit & 9 kW &- \\
         Initial grid voltage at PCC & SG & 400 V &1.0 \\
         Initial grid frequency at PCC & ASM & 50 Hz &0.5 \\
         Initial active power FC 1 & FC 1 & 3 kW &0.3 \\
         Initial active power FC 2 \tablefootnote{A STATCOM usually does not provide steady state power. However, to realize a negative $\Delta$p, positive steady-state power must be provided by the converter at the test bed. In the data evaluation, this steady state power is subtracted.} 
         & FC 2 & 1 kW &0.1 \\
         Initial active power FC 3 & FC 3 & 2 kW &0.2 \\
         \hline
         Data log resolution & PLC & 1 kHz &- \\
         DVPP task tick rate (sample time) & PLC & 10 ms&- \\
         UDP task tick rate (sample time) & PLC & 10 ms&- \\
         Grid frequency measurement resolution & PLC & 1 mHz&- \\
         Grid frequency measurement update rate & PLC & 6,67 ms&- \\
    \bottomrule
    \end{tabular}
        \vspace{-3mm}
    \end{minipage}
\label{tab:testbed_para}
\end{table}
\renewcommand{\arraystretch}{1}\normalsize

\subsection{Case Study I}
\begin{figure}[t!]
        \centering
        \scalebox{0.5}{\includegraphics[]{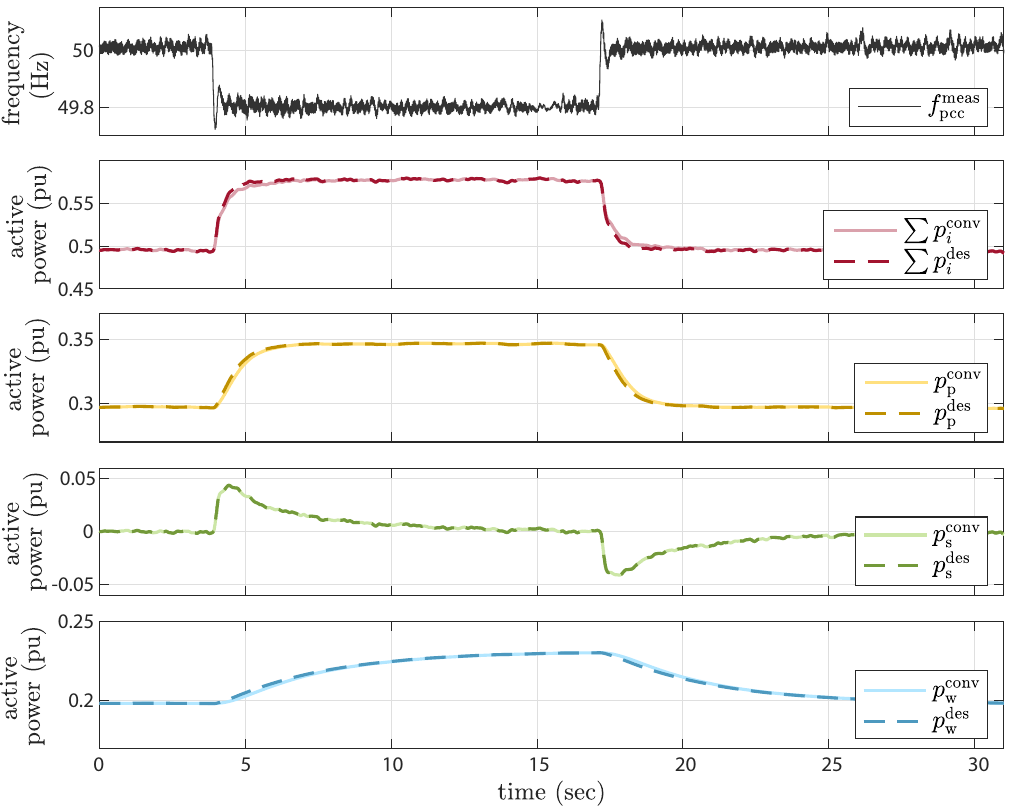}}
        \vspace{-3mm}
        \caption{Active power response of the DVPP based on DPFs during a frequency jump of $\pm$ 200 mHz in case study I. The dashed lines indicate the desired active power response behavior.}
                \vspace{-3mm}
        \label{fig:DPF_response_FJ}
    \end{figure}

To investigate the dynamic active power response behavior of the DVPP during long-term changes in the grid-frequency, we employ a $\pm$ 200 mHz frequency jump of the grid-frequency by adjusting $f_\mathrm{grid}^\mathrm{ref}$ of the ASM SG, accordingly. In doing so, the speed control of the ASM SG is executed in the standard mode to ensure a tight tracking of the step change $f_\mathrm{grid}^\mathrm{ref}$, independent of any power imbalances in the grid. The experimental results are illustrated in Fig. \ref{fig:DPF_response_FJ}. It becomes apparent how the aggregated DVPP exhibits an accurate matching of the desired active power injection (dashed lines). In particular, the individual DVPP units, i.e., the wind power system, the PV system and the STATCOM accomplish an active power injection behavior according to their specified DPF characteristic as a low-pass and band-pass filter behavior in \eqref{eq:DPF_wind_pv} and \eqref{eq:DPF_statcom}, respectively, which has been carefully selected to account for the heterogeneous time scales of the DER dynamics as well as the individual steady-state power capacity limitations. 
 \begin{figure}[t!]
        \centering
        \scalebox{0.51}{\includegraphics[]{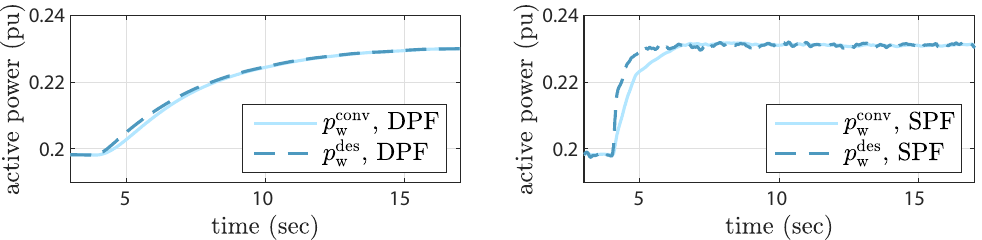}}
        \vspace{-3mm}
        \caption{Active power response of the wind power plant during a frequency jump of 200 mHz when using an SPF in comparison to a DPF in case study I.}
        \vspace{-3mm}
        \label{fig:wind_SPF_DPF_response_FJ}
    \end{figure}
    \begin{figure}[t!]
        \centering
        \scalebox{0.51}{\includegraphics[]{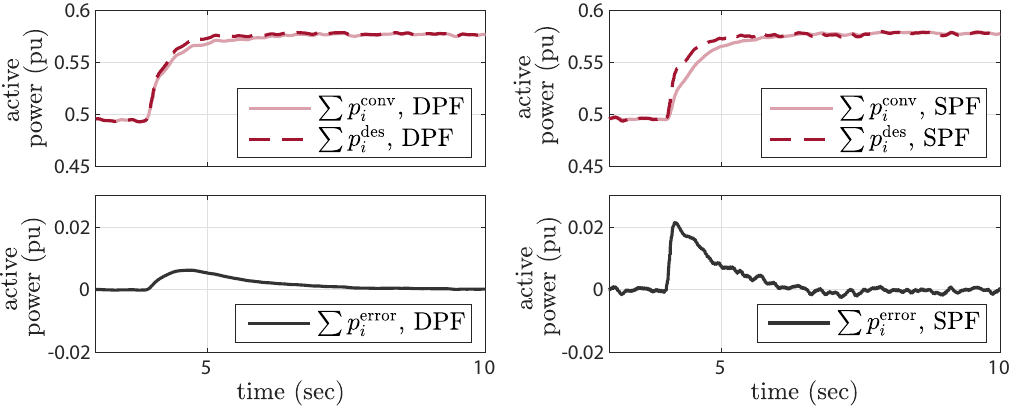}}
        \vspace{-7mm}
        \caption{Total active power response of the DVPP during a frequency jump of 200 mHz when using SPFs in comparison to DPFs in case study I.}
        \label{fig:DVPP_SPF_DPF_response_FJ}
                        \vspace{-3mm}
    \end{figure}

As an example, if one was to assign a \textit{static} participation factor (SPF) $m_\mathrm{w}(s)=\mu_\mathrm{w}$ to the wind generation system, the associated, more aggressive desired response behavior cannot be provided within the local limits of the wind turbine system (Fig. \ref{fig:wind_SPF_DPF_response_FJ}). This is in contrast to the proposed concept of DPFs, which allows to select the desired response behavior of the wind turbine to be sufficiently slow, such that the local time scales can be taken into account. Moreover, as a side benefit of a less aggressive wind turbine response behavior, the wear-out of the wind generation system components can be reduced.

The aggregated active power response behavior of a DVPP fully based on SPFs is depicted in Fig. \ref{fig:DVPP_SPF_DPF_response_FJ}. Following the same reasoning as for the previous wind turbine example, it becomes apparent how the total active power response of the DVPP achieves a distinctively less accurate matching of the desired behavior than the proposed DVPP based on DPFs. Notice that, for the sake of consistency, we have kept the same dc gains for the SPFs as for the DPFs in \eqref{eq:DPF_wind_pv} and \eqref{eq:DPF_statcom}, i.e., $m_\mathrm{w}(s)=\mu_\mathrm{w}$, $m_\mathrm{p}(s)=\mu_\mathrm{p}$ and $m_\mathrm{s}(s)=0$. However, alternatively, one could also use different dc gains and thus also allocate some steady-state contribution to the STATCOM. If so, one would observe how the STATCOM collapses after a short time of providing steady-state power caused by its limitation in energy and the resulting inability to provide regulation on long-time scales (not shown in our experimental results of this paper).

Finally, the differences between the different disaggregation strategies of the DVPP, i.e., SPFs vs. DPFs, can be even more severe in larger system topologies and/or for other types of DER aggregations, see \cite{haeberle2021control} for illustrative examples.



    \begin{figure}[t!]
        \centering
                        
        \scalebox{0.5}{\includegraphics[]{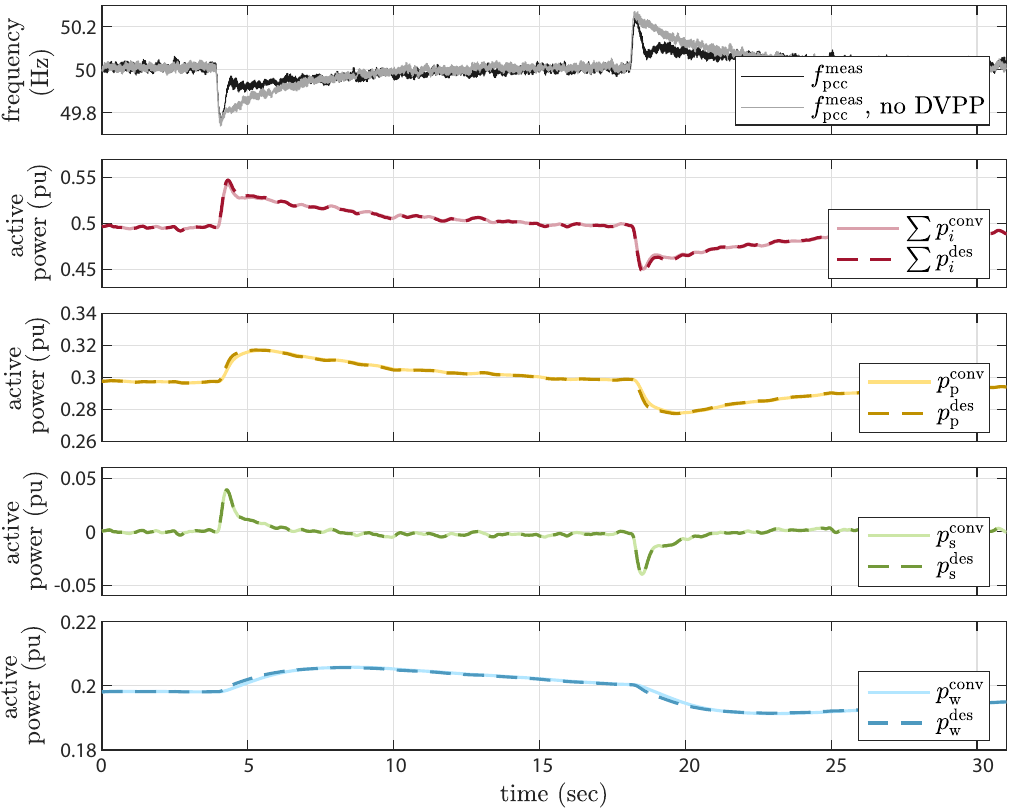}}
                \vspace{-3mm}
        \caption{Active power response of the DVPP during a $\pm$ 1kW load change in case study II.}
        \label{fig:DPF_response_LJ}
        \vspace{-3mm}
    \end{figure}
    
\subsection{Case study II}
Next, we investigate the impact of the DVPP active power response on the grid frequency during a $\pm$ 1kW load jump of the load unit by means of a stepwise load increase followed by a stepwise load reduction. To do so, the speed control of the ASM SG is now executed in the slow mode, which allows for some manipulable flexibility of the grid frequency dynamics during a short time interval up to approximately 10 seconds before the PI-controlled speed regulation of the ASM SG has settled. The experimental results are shown in Fig. \ref{fig:DPF_response_LJ}. We can see how the dynamic active power response of the aggregated DVPP exhibits an accurate matching of the desired active power injection (dashed lines). In particular, similar to case study I before, the individual DVPP units contribute to the overall active power injection according to their selected DPF characteristic in \eqref{eq:DPF_wind_pv} and \eqref{eq:DPF_statcom}, respectively. 

However, in contrast to case study I where the ASM SG was tightly tracking the imposed frequency jump, we can now observe how the overall grid-frequency response is affected by the dynamic DVPP control. In particular, by comparing the frequency response during the load jump with and without the DVPP control (see Fig. \ref{fig:DPF_response_LJ}, top), it becomes apparent how the additional dynamic active power injection of the DVPP has a supportive impact in bringing the grid frequency back to its nominal value faster, while additionally slightly reducing the frequency nadir. Nevertheless, it has to be noted that we observed some undesirable interactions of the dominant PI-based ASM
speed control when the desired DVPP response was specified inappropriately to the test bed dynamics (not shown), such that our experiments are
limited to a sufficient DVPP injection that dynamically matched the test bed. However, in reality, the grid-frequency would not be exposed to such a dominant PI speed-control of other synchronous generators during fast dynamic transients, such that we would not face such undesired DVPP interactions in a realistic setup.


\section{Conclusion}\label{sec:conclusion}
We have presented an experimental validation of the DVPP control concept in \cite{haeberle2021control} on a multi-converter PHIL test bed system. The DVPP was composed of a wind generation system, a PV system, and a STATCOM with small storage to collectively provide grid-following fast frequency regulation during grid-frequency changes and load variations. We experimentally validated the DVPP control concept and particularly demonstrated the benefit of the proposed DPFs over conventional static allocation schemes. Future work should consider further experimental investigations of the DVPP performance during time-varying generation capacities of the DVPP units, where the latter are equipped with adaptive DPFs.

\section{Acknowledgements}
This paper is based upon work supported by the European Union's Horizon 2020 research and innovation programme under Grant Agreement Number 883985.

\end{document}

%% file: Figures/DVPP_scheme.tex
\resizebox{0.4\textwidth}{!}{

\begin{tikzpicture}[scale=1, every node/.style={scale=1.1}]
\draw [rounded corners = 3, dashed,black!60,fill=black!5] (-3.8,3.5) rectangle (-1.4,-0.3);
\draw  [rounded corners = 3, dashed,black!60,fill=black!5]  (1.9,2.3) rectangle (4.1,0.9);
\draw [fill=wind!50](-3.5,3.2) rectangle (-1.7,2.4);
\node at (-2.6,2.8) {wind};

\draw [fill=PV!50] (-3.5,2) rectangle (-1.7,1.2);
\node at (-2.6,1.6) {PV};

\draw  [fill=STATCOM!50](-3.5,0.8) rectangle (-1.7,0);
\node [scale=0.9] at (-2.6,0.4) {STATCOM};
\draw (-1.7,2.8) -- (-1,2.8); 
\draw (-1.7,1.6) -- (-1,1.6) node (v1) {}; 
\draw (-1.7,0.4) -- (-1,0.4); 
\draw[ultra thick] (-1,3.2) -- (-1,0);
\node [scale = 0.9] at (-0.9,3.5) {PCC};

\node [scale=0.9,black!60] at (-2.6,3.7) {DVPP};
\draw (3.8,1.6) -- (5,1.6) node (v2) {};
\draw  plot[smooth, tension=.7] coordinates {(6.2,3.5) (5.1,2.7) (v2) (5.2,0.8) (5.7,0.1) (6.2,-0.3)};

\draw (-1,1.6) -- (-0.5,1.6) node (v2) {};
\draw  plot[smooth, tension=.7] coordinates {(0.7,3.5) (-0.4,2.7) (v2) (-0.3,0.8) (0.2,0.1) (0.7,-0.3)};
\node at (5.7,1.9) {power};
\node at (5.7,1.5) {grid};
\node at (0.2,1.9) {power};
\node at (0.2,1.5) {grid};
\draw  (2.2,2) rectangle (3.8,1.2);
\node [scale=1.1]at (3,1.6) {$T_\mathrm{des}(s)$};

\draw [ultra thick](4.5,2.2) -- (4.5,1);
\node [scale=0.9]at (4.5,2.5) {PCC};
\node [scale=1.2]at (1.3,1.6) {$\approx$};
\end{tikzpicture}

}

%% file: Figures/DVPP_control_setup.tex
\resizebox{0.4\textwidth}{!}{
\tikzstyle{roundnode}=[circle,draw=black!60,fill=black!5,scale=0.75]
\begin{tikzpicture}[scale = 1, every node/.style={scale=1}]
\draw [dashed, rounded corners =3,black!60, fill=black!5] (-4.3,4.6) rectangle (3,-3.9);

\draw  [fill=STATCOM!20](1.7,-1.7) rectangle (-3,-3.3);
\draw [rounded corners = 3,fill=black!40] (-2.7,-2.1) rectangle (-1.3,-2.7);
\node at (-2,-2.4) {control};
\draw [fill=STATCOM!60] (-0.9,-1.9) rectangle (1.1,-2.9);
\node at (0.1,-2.4) {STATCOM};
\draw[-latex] (-1.3,-2.4) -- (-0.9,-2.4);
\draw[-latex] (1.1,-2.4) -- (1.4,-2.4) -- (1.4,-3.1) -- (-2,-3.1) -- (-2,-2.7); 

\draw  [fill=wind!20](1.7,3.9) rectangle (-3,2.3);
\draw [rounded corners = 3,fill=black!40] (-2.7,3.5) rectangle (-1.3,2.9);
\node at (-2,3.2) {control};
\draw [fill=wind!60] (-0.9,3.7) rectangle (1.1,2.7);
\node at (0.1,3.4) {wind power};
\node at (0.1,3) {plant};
\draw[-latex] (-1.3,3.2) -- (-0.9,3.2);
\draw[-latex] (1.1,3.2) -- (1.4,3.2) -- (1.4,2.5) -- (-2,2.5) -- (-2,2.9); 

\draw  [fill=PV!20](1.7,1.1) rectangle (-3,-0.5);
\draw [rounded corners = 3,fill=black!40] (-2.7,0.7) rectangle (-1.3,0.1);
\node at (-2,0.4) {control};
\draw [fill=PV!60] (-0.9,0.9) rectangle (1.1,-0.1);
\node at (0.1,0.6) {PV power};
\node at (0.1,0.2) {plant};
\draw[-latex] (-1.3,0.4) -- (-0.9,0.4);
\draw[-latex] (1.1,0.4) -- (1.4,0.4) -- (1.4,-0.3) -- (-2,-0.3) -- (-2,0.1);

\draw [-latex] (1.4,3.2) -- (2.6,3.2) -- (2.6,0.55);
\draw[-latex] (-3.9,0.4) -- (-3.9,3.2) -- (-2.7,3.2);
\draw[-latex] (-5.5,0.4) -- (-2.7,0.4);
\draw [-latex] (-3.9,0.4) -- (-3.9,-2.4) -- (-2.7,-2.4);
\node at (-0.6,2.05) {wind generation system};
\node at (-0.6,-0.75) {PV generation system};
\node  at (-0.6,-3.55) {STATCOM system};

\fill [black] (-3.9,0.4) circle (0.4mm); 
\fill[black] (1.4,3.2) circle (0.4mm); 
\fill [black] (1.4,0.4)circle (0.4mm); 
\fill [black] (1.4,-2.4) circle (0.4mm); 
\node at (-0.4,0.7) {};

\draw [-latex](1.2,0.4) -- (2.45,0.4); 
\draw [-latex](1.2,-2.4) -- (2.6,-2.4) -- (2.6,0.25); 
\draw[-latex] (2.7,0.4) -- (4.1,0.4);
\node [roundnode] at (2.6,0.4) {};

\node [black!60,scale=0.9] at (-0.6,4.8) {DVPP};

\draw [dashed, rounded corners = 3, black!60, fill=black!5] (-4.3,-4.9) rectangle (3,-6.1);
\draw  (-1.7,-5.1) rectangle (0.5,-5.9);
\node [scale=1.1] at (-0.6,-5.5) {$T_\mathrm{des}(s)$};
\draw[-latex] (-5.5,-5.5) -- (-1.7,-5.5); 
\draw [-latex](0.5,-5.5) -- (4.1,-5.5);
\node [scale=1.2] at (-0.6,-4.3) {$\approx$};
\node[black!60] at (-0.6,-4.7) {desired aggregate behavior};
\node at (2.8,0.8) {$+$};
\node at (2.1,0.2) {$+$};
\node at (2.8,0) {$+$};
\node at (-4.9,0.7) {$\Delta f_\mathrm{pcc}$};
\node at (3.6,0.7) {$\Delta p_\mathrm{pcc}$};
\node at (3.6,-5.2) {$\Delta p_\mathrm{pcc}$};
\node at (-4.9,-5.2) {$\Delta f_\mathrm{pcc}$};
\node at (2.1,3.5) {$\Delta p_\mathrm{w}$};
\node at (2.1,0.7) {$\Delta p_\mathrm{p}$};
\node at (2.1,-2.1) {$\Delta p_\mathrm{s}$};
\node at (-0.6,4.15) {$T_\mathrm{w}(s)$};
\node at (-0.6,1.35) {$T_\mathrm{p}(s)$};
\node at (-0.6,-1.45) {$T_\mathrm{s}(s)$};
\end{tikzpicture}
}

%% file: Figures/DVPP_TF_scheme.tex
\resizebox{0.48\textwidth}{!}{
\tikzstyle{roundnode}=[circle,draw=black!60,fill=black!5,scale=0.75]

\begin{tikzpicture}[scale = 1, every node/.style={scale=1.5}]
\draw[dashed, rounded corners =3,black!60, fill=green!5] (10.5,1) rectangle (23,-9);
\draw[dashed, rounded corners =3,black!60, fill=red!5] (8.5,1) -- (10.5,1) -- (10.5,-9) -- (23,-9) -- (23,1) -- (26,1) -- (26,-13.5) -- (8.5,-13.5) -- (8.5,1) ;

\draw[fill=white] (11,0.5) rectangle (13,-8);

\draw (9.5,-0.5)[-latex]  -- (11,-0.5);

\node [align=center] at (9.5,0) {$p^\mathrm{ref}_{i}$};
\node [align=center] at (12,-4) {desired \\ local\\ DVPP\\ be-\\haviors\\ (Sec.\ref{subsec:DPF_struct})};
\draw[rounded corners = 3,fill=black!40] (15,0.5) rectangle (17,-1.5);
\draw[fill=PV!50] (19,0.5) rectangle (21,-1.5);

\draw (13,-0.5) -- (13.85,-0.5)  (14.15,-0.5) -- (15,-0.5)  (17,-0.5) -- (19,-0.5)  (21,-0.5) -- (24,-0.5)[-latex] -- (24,-3.85);
\node [align=left] at (14,0) {$\Delta p^\mathrm{des}_\mathrm{p}$};
\node [align=left] at (18,0) {$\Delta p^\mathrm{ctrl}_\mathrm{p}$};
\node [align=left] at (22,0) {$\Delta p^\mathrm{conv}_\mathrm{p}$};

\draw (22,-0.5) -- (22,-2) -- (14,-2)[-latex] -- (14,-0.65);
\draw (22,-0.5)[fill=black] circle (0.05);

\draw (14,-0.5) circle (0.15) node [anchor=north east]{$-$};

\node [align=center] at (16,-0.5) {${C}_\mathrm{p}(z)$};
\node [align=center] at (20,-0.5) {${P}_\mathrm{p}(z)$};

\draw[fill=STATCOM!50] (19,-3)	rectangle (21,-5);

\draw (13,-4) -- (19,-4)  (21,-4)[-latex] -- (23.85,-4);
\node [align=left] at (14,-3.5) {$\Delta p^\mathrm{des}_\mathrm{s}$};
\node [align=left] at (22,-3.5) {$\Delta p^\mathrm{conv}_\mathrm{s}$};

\node [align=center] at (20,-4) {${P}_\mathrm{s}(z)$};

\draw[rounded corners = 3,fill=black!40] (15,-6) rectangle (17,-8);
\draw[fill=wind!50] (19,-6) rectangle (21,-8);

\draw (13,-7) -- (13.85,-7)  (14.15,-7) -- (15,-7)  (17,-7) -- (19,-7)
	 (21,-7) -- (24,-7)[-latex] -- (24,-4.15);
\node [align=left] at (14,-6.5) {$\Delta p^\mathrm{des}_\mathrm{w}$};
\node [align=left] at (18,-6.5) {$\Delta p^\mathrm{ctrl}_\mathrm{w}$};
\node [align=left] at (22,-6.5) {$\Delta p^\mathrm{conv}_\mathrm{w}$};

\draw (22,-7) -- (22,-8.5) -- (14,-8.5)[-latex] -- (14,-7.15);
\draw (22,-7)[fill=black] circle (0.05);

\draw (14,-7) circle (0.15) node [anchor=north east]{$-$};https://de.overleaf.com/project/6445822ec80a9ca6f2b04884

\node [align=center] at (16,-7) {${C}_\mathrm{w}(z)$};
\node [align=center] at (20,-7) {${P}_\mathrm{w}(z)$};

\draw (24,-4) circle (0.15);

\draw[fill=white] (15,-10) rectangle (21,-12);

\draw (24.15,-4) -- (25,-4)	-- (25,-11)[-latex] -- (21,-11);

\draw (15,-11) -- (9.5,-11) -- (9.5,-4)[-latex]  -- (11,-4);

\draw (19,-13) -- (18,-13)[-latex]  -- (18,-12);

\node [align=center] at (18,-11) {equivalent grid dynamics \\(Sec.\ref{subsec:equi_grid_eq})};
\node [align=center] at (9.5,-3.5) {$\Delta f^\mathrm{meas}_\mathrm{pcc}$};
\node [align=center] at (25,-3.5) {$\Delta p_\mathrm{pcc}$};
\node [align=left] at (21,-13) {$ \textstyle\sum_{j} p^\mathrm{load}_{j}, \omega^\mathrm{ref}_\mathrm{grid}$};


\end{tikzpicture}

}

%% file: full_paper_main.bbl
\section{References}
{\small
\begin{thebibliography}{4}
\bibitem{marinescu2022dynamic} Marinescu, B., Gomis-Bellmunt, O., Dörfler, F., et al.: `Dynamic virtual power plant: A new concept for grid integration of renewable energy sources', IEEE Access, 2022, vol. 10, pp. 104980-104995.

\bibitem{haeberle2021control} Häberle, V., Fisher, M. W., Araujo, E. P., et al: `Control design of dynamic virtual power plants: An adaptive divide-and-conquer approach', IEEE Transactions on Power Systems, 2021.

\bibitem{bjork2022dynamic} Björk, J., Johansson, K. H., and Dörfler, F.:`Dynamic virtual power plant design for fast frequency reserves: Coordinating hydro and wind', IEEE Transactions on Control of Network Systems, 2022.

\bibitem{zhong2021coordinated} Zhong, W., Chen, J., Liu, M., et al: `Coordinated control of virtual power plants to improve power system short-term dynamics', Energies, 2021, vol. 14, no. 4, p. 1182.

\bibitem{AlbernazLacerda2022} Albernaz Lacerda, V. and Girona-Badia, J. and Prieto-Araujo, E. and Gomis-Bellmunt, O. and Kusche, S. and Pöschke, F. and Schulte, H.:`Modelling Approaches of Power Systems Considering Grid-Connected Converters and Renewable Generation Dynamics', European Conference on Power Electronics and Applications. Hannover: 2022, S. 1-7. 

\bibitem{Poeschke2020} Pöschke, F., Gauterin, E., Kühn, M., Fortmann, J., and Schulte, H., “Load mitigation and power tracking capability for wind turbines using
linear matrix inequality-based control design,” Wind Energy, 2020.

\bibitem{Poeschke2021} Pöschke, F. and Schulte, H., “Evaluation of different APC operating strategies considering turbine loading and power dynamics for grid support,” Wind Energy Science Discussions, vol. 2021, pp. 1–14, 2021. 


\end{thebibliography}
}
